\documentclass{ecai2010}
\usepackage{times}
\usepackage{latexsym}
\usepackage{amssymb}
\usepackage{amsmath}
\usepackage{tikz}
\usepackage{helvet,graphicx,multicol}
\usepackage{subfigure}

\newcommand{\G}{\mathcal{G}}
\newcommand{\J}{\mathcal{J}}
\newcommand{\T}{\mathcal{T}}

\newcommand{\temp}{\delta}
\newcommand{\topo}[1]{{#1}}
\newcommand{\fast}{\hat{\delta}}

\newcommand{\ecc}{\varepsilon}
\newcommand{\TVG}{\ensuremath{\G=(V,E,\T,\rho,\zeta)}}
\newcommand{\TVGL}{\ensuremath{\G=(V,E,\T,\rho)}}

\begin{document}

\title{Time-Varying Graphs and Social Network Analysis: Temporal
Indicators and Metrics}

\author{Nicola Santoro\institute{Carleton University, Ottawa, Canada, E-mail: santoro@scs.carleton.ca}~,~
Walter Quattrociocchi\institute{University of Siena, Italy, E-mail: walter.quattrociocchi@unisi.it}~,~
Paola Flocchini$^3$,~
Arnaud Casteigts\institute{University of Ottawa, Canada, E-mail:\{casteig,flocchin\}@site.uottawa.ca}~,~ and 
Frederic Amblard\institute{IRIT-CNRS, Toulouse, France, E-mail:{frederic.amblard@univ-tlse1.fr}} 
}

\maketitle

\begin{abstract}
Most instruments -~formalisms, concepts, and metrics~- for social networks analysis fail to capture their dynamics. Typical systems exhibit different scales of dynamics, ranging from the fine-grain dynamics of interactions (which recently led researchers to consider {\em temporal} versions of {\em distance}, {\em connectivity}, and related indicators), to the evolution of network properties over longer periods of time. 
This paper proposes a general approach to study that evolution for both {\em atemporal} and {\em temporal} indicators, based respectively on sequences of static graphs and sequences of time-varying graphs that cover successive time-windows. All the concepts and indicators, some of which are new, are expressed using a {\em time-varying graph} formalism recently proposed in~\cite{CFQS2010a}.
\end{abstract}

\section{Introduction}

Social networks have drawn a lot of attention in the past few years, and the analysis of their dynamics represents a pressing scientific challenge. The research efforts in this area strive to understand   the driving forces behind the evolution of social networks and their articulations within
 social dynamics, e.g., opinion dynamics, the epidemic or innovation diffusion, 
 the teams formation and so on   (\cite{amblard01,Moore2000,Lelarge09,Carley02,Powell05,Guimera05,QuattrociocchiPC09,castellano07,quattrociocchi2010e}). In other words, it is known that individuals are influenced ({\it e.g.} concerning their opinion) through their social network, it is also known that individuals take into account others' attributes when deciding to evolve their social network, but yet qualitatively not much is known concerning the dynamical patterns that are produced by such an interplay.

Curiously enough, everybody agrees on the stance that social networks are dynamic, e.g. individuals join, participate, attract, compete, cooperate, disappear, and affect the shape and strength of the network and its relationships.
Yet, the current instruments (definitions, models, metrics) are mainly drawn for static networks and generally fail to capture the evolution of phenomena and their dynamical properties -- {\em temporal dimension} -- focusing instead on structural \cite{KeK02} or statistical aspects \cite{Snijders2001} of the systems.  
As stated in \cite{LESK10}, the central problem in this area is the definition of mathematical models able to capture and to reproduce properties observed on the real networks. 

The increasing availability of real datasets ({\it e.g.} e-mails logs, online forums, or meta-data on scientific publishing), as well as development of smartphones, vehicular networks, and satellite networks have recently fostered research on dynamic networks and caused the appearance of new dedicated concepts. In particular, early works around transportation and {\em delay-tolerant} networks (those networks characterized by an absence of instant end-to-end connectivity) have led to the concept of {\em journey}~\cite{BFJ03} -~also called {\em schedule-conforming path}~\cite{Berman96}, {\em time-respecting path}~\cite{Holme05,KKK00}, or {\em temporal path}~\cite{ChMMD08,TSM+09}. Journeys can be seen as a particular kind of path whose edges do not necessarily follow one another instantly, but instead induces waiting times at intermediate nodes. 

A direct consequence of considering {\em journeys} instead of {\em paths} is that all the concepts usually built on top of paths can in turn take a temporal meaning. This includes the concept of {\em temporal distance}~\cite{BFJ03} -~also called {\em reachability time}~\cite{Holme05}, {\em information latency}~\cite{KosKW08}, or {\em temporal proximity}~\cite{Kostakos09}~-, which accounts for the minimal speed of information propagation between two nodes, and the concept of {\em temporal connectivity}~\cite{BF03} based on the existence of journeys. On the social network side, recent studies focused on measuring the temporal distance between individuals based on e-mail datasets~\cite{KosKW08, Kostakos09} or inter-meeting times~\cite{TSM+09}. Very recently, {\em temporal betweenness} and {\em temporal closeness} were also considered in a social network context in~\cite{TMMLN10}. All these {\em temporal} indicators complete the set of {\em atemporal} indicators usually considered in social network analysis, such as (the usual versions of) distance and diameter, density, clustering coefficient, or modularity, to name a few. It is important to keep in mind that these indicators, whether temporal or atemporal, essentially accounts for network properties at a reasonably short time-scale ({\em fine-grain} dynamics). They do not reflect how these properties evolve over longer periods of time ({\em coarse-grain} dynamics).

In this paper, we propose a general approach to look at the evolution of both atemporal and temporal indicators. Looking at the evolution of atemporal indicators can be done by representing the evolution of the network as a sequence of {\em static} graphs, each of which represents the aggregated interactions over a given time-window. Atemporal indicators can then be normally measured on these graphs and their evolution studied over time. The case of temporal indicators is more complex because the corresponding evaluation cannot be done on static graphs. The proposed solution is therefore to look at the evolution of temporal indicators through a {\em sequence} of shorter {\em time-varying graphs}, which are {\em temporal subgraphs} of the original time-varying graph, covering successive time-windows. We discuss several examples of indicators, both temporal and atemporal, some of which are new. The evolution of some atemporal indicators is accompanied with recent experimental results from~\cite{AQ2010a}, based on online data on scientific networking consisting of dated co-authoring and citation records.
 We first present the time-varying graph (TVG) formalism from~\cite{CFQS2010a}, which we use to express all temporal concepts and evolution properties in a concise and elegant way. We then discuss the two suggested approaches to study  the evolution of {\em atemporal} and {\em temporal} indicators, respectively.

\section{Dynamic Networks as Time-Varying Graphs}
\label{sec:Model}

This section presents the {\em time-varying graph} formalism (TVG) recently introduced in~\cite{CFQS2010a}. This formalism is semantically equivalent to other graph formalisms, 
like that of {\em evolving graphs}~\cite{Fer04}, but suggests in comparison 
an {\em interaction-centric} point of view. This point of view was also present in the  
time-labelling function of~\cite{KKK00}, but only for punctual contacts and latencies.
The TVG formalism allows us a concise and elegant formulation of temporal concepts and properties.

\subsection{The TVG Formalism}
 
Consider a set of entities $V$ (or {\em nodes}), a set of relations $E$ between these entities ({\em edges}), and an alphabet $L$ accounting for any property such a relation could have ({\em labels}); that is, $E \subseteq V \times V \times L$. The definition of $L$ is domain-specific, and therefore left open --a label could represent for instance a particular type of relation in a social network, a type of carrier in a transportation networks, or a communication medium in communication networks. For generality, $L$ is assumed to possibly contain multi-valued elements (e.g. $<${\it satellite link; bandwidth of 4\,MHz; encryption available;...}$>$ ). The set $E$ enables multiple relations between a given pair of entities, as long as these relations have different properties, that is, for any $e_1=(x_1,y_1,\lambda_1)\in E,e_2=(x_2,y_2,\lambda_2) \in E$, $(x_1=x_2 \wedge y_1=y_2 \wedge \lambda_1=\lambda_2)\implies e_1=e_2$.

The relations between entities are assumed to take place over a time span $\T \subseteq \mathbb{T}$ called the {\em lifetime} of the system. The temporal domain $\mathbb{T}$ is generally assumed to be $\mathbb{N}$ for discrete-time systems or $\mathbb{R}$ for continuous-time systems. We denote by time-varying graph the structure \TVG, where
$\rho: E \times \T \rightarrow \{0,1\}$, called {\em presence function}, indicates whether a given edge is present at a given time, and
$\zeta: E \times \T \rightarrow \mathbb{T}$, called {\em latency function}, indicates the time it takes to cross a given edge if starting at a given date.

Such a formalism can arguably describe a multitude of different scenarios, including:

\begin{itemize}
  \item Transportation networks - e.g. aviation, where nodes are the cities, directed edges are regular flights, whose departure dates are given by {\em punctual} presences, and flight duration by non-nil latencies.
  \item Communication networks - e.g. wireless mobile networks, where an edge is present whenever its two endpoints are within range, the latency corresponding here to the time to propagate a message.
  \item Complex systems, among which social networks - e.g. scientific networks, where the nodes are scientists, and the edges (possibly both directed and undirected) account for example for citations or collaborations.
 \end{itemize}

These examples illustrate the spectrum of models over which the TVG formalism can stretch. As observed, some contexts are intrisically simpler than others and call for restrictions (e.g. directed vs. undirected edges, single vs. multiple edges, punctual vs. lasting relations). Further restrictions may apply. For example the latency function could be decided constant over time, over the edges,  over both, or simply ignored. In fact, a vast majority of work in social networks does not require such information (e.g.,  the propagation time of an email is of little interest to the understanding of a community behavior). Since the scope of this paper is social network analysis, we will deliberately omit the latency function and consider TVGs described as \TVGL.

\subsection{Journeys and related Temporal Concepts}
\label{sec:journeys}

A crucial concept in time-varying graphs is that of {\em journey} which is the temporal extension of the notion of path, and forms the basis of most recently introduced temporal concepts. A sequence of couples $\J=\{(e_1,t_1),$ $(e_2,t_2) \dots,$ $(e_k,t_k)\}$, such that $\{e_1, e_2,...,e_k\}$ is a  walk in $G$,
 is a {\em journey} in $\G$ if and only if $\forall i,  1\leq i < k$, $\rho(e_i,t_i)=1$ and $t_{i+1}\ge t_i$. 
 We denote by $departure(\J)$, and $arrival(\J)$, the starting date $t_1$ and the last date $t_k$ of a journey $\J$, respectively.
 Journeys can be thought of as {\em paths over time} from a source to a destination and therefore have both a {\em topological} and a {\em temporal} length.
 The {\em topological length} of $\J$ is the number $\topo{|\J|}= k$ of couples in $\J$ (i.e., the number of {\em hops}); its {\em temporal length} is its end-to-end duration:  $\topo{||\J||}= arrival(\J) - departure(\J)$.

Let us denote by $\J^*$ the set of all possible journeys in a time-varying graph $\G$, and by  $\J^{*}(u,v) \subseteq \J^*$ those journeys starting at node $u$ and ending at node $v$. 
In a time-varying graph, there are three natural distinct measures of {\em distance}, and thus three different types of ``minimal" journeys.

\begin{itemize}
\item The {\em shortest distance}  from a node $u$ to a node $v$ at time $t$ is simply
  $\topo{d}^{t}(u,v) = Min\{\topo{|\J|}:\J \in \J^{*}(u,v) \wedge departure(\J) \ge t \}$.
     
\item  The {\em foremost distance} from $u$ to $v$ at time $t$ is 
  $\temp^{t}(u,v) = Min\{arrival(\J)- t: \J \in \J^{*}(u,v) \wedge departure(\J)\ge t\}$.

\item    The {\em fastest distance} from $u$ to $v$ at time $t$ is defined as 
    $\fast^{t}(u,v)  = Min\{\topo{||\J||}:\J \in \J^{*}(u,v) \wedge departure(\J) \ge t \}$.
\end{itemize}

A journey
   $\J \in\J^{*}(u,v)$  
   with  $ departure(\J) \ge t $ 
is said to be  {\em shortest} at time $t$ if    $|\J|= \temp^{t}(u,v)$;
{\em foremost} at time $t$ if    $arrival(\J)- t =\temp^{t}(u,v)$; and  {\em fastest} 
at time $t$ if $||\J|| =\fast^{t}(u,v)$.

Whether in the contexts of social networks or communication networks, a number of higher concepts have been recently defined on top of these. They include new meanings of {\em connectivity} and {\em connected components}~\cite{BF03}, {\em temporal eccentricity} and {\em temporal diameter}~\cite{BFJ03}, or {\em temporal betweenness} and {\em temporal closeness}~\cite{TMMLN10}, among others.
As discussed in the introduction, these concepts allow for novel insights on the way nodes interact at a small time-scale ({\em fine-grained} dynamics), but do not reflect the way the network {\em evolves} at over longer periods of time ({\em coarse-grain} dynamics).

\section{Capturing the Evolution}\label{sec:evolution}


In this section we introduce a framework to study the behavior  of network parameters 
(or indicators) during the lifetime of a time-varying graph.  
Two types of indicators are described: {\em atemporal} and {\em temporal} ones. 
Atemporal parameters are defined on static networks and their evolution 
over time can  be observed by measuring them over sequences of static graphs, where each graph of the sequence corresponds to the aggregation of interactions that occur in a given interval of time (we call them {\em footprints} of a TVG).
Temporal indicators, on the other hand, are only defined on time-varying graphs,
taking into account their temporal nature. The evolution of such indicators requires to consider a sequence of (non-aggregated) time-varying graphs, each of which corresponds to a {\em temporal subgraph} of the original one for the considered interval.

%
%
 
\subsection{Evolution of Atemporal  Indicators} \label{static}
\label{sec:atemporal}

\subsubsection{Methodological approach}

\paragraph{TVGs as a sequence of footprints.}

Given a TVG \TVGL, one can define the {\em footprint} of this graph from $t_1$ to $t_2$ as the static graph $G^{[t_1,t_2)}=(V,E^{[t_1,t_2)})$ such that $\forall e \in E, e \in E^{[t_1,t_2)} \iff \exists t \in [t_1,t_2), \rho(e,t)=1$. In other words, the footprint aggregates interactions over a given time window into static graphs. Let the lifetime  $\T$ of the time-varying graph be partitioned in consecutive sub-intervals $\tau = [t_0,t_1), [t_1,t_2) \ldots [t_i,t_{i+1}), \ldots$; where each $[t_k,t_{k+1})$ can be noted $\tau_k$. We call {\em sequence of footprints} of $\G$ according to $\tau$ the sequence {\em SF}$(\tau) = G^{\tau_0},G^{\tau_1}, \ldots$.
Considering this sequence with a sufficient size of the intervals allows to overcome the strong fluctuations of fine-grain interactions, and focus instead on more general trends of evolution. Note that the same approach could be considered with a sequence of intervals that are {\em overlapping} ({\it i.e.,} a sliding time-window) instead of disjoint ones. Another axis of variation can be considered whether or not the set of nodes in each $G^{\tau_i}$ is also varying, e.g. being restricted to nodes that have at least one adjacent edge in $E^{\tau i}$ (which is the case in the experimental results shown below).

\paragraph{Looking at atemporal parameters.}
Since every graph in $SF$ is static,  any classical network parameter (degree, neighborhood, density, diameter, modularity, {\it etc.}) can be directly 
measured on it. 
When observing the evolution of  a parameter over SF, one can achieve different levels of granularity by varying the size of the footprint intervals. Depending on the parameter and on the application, different choices of 
granularity are more appropriate to capture a meaningful behavior. At one extreme, each interval could correspond to the smallest time unit (in discrete-time systems), or to the time between any two consecutive modification of the graph. In these cases every footprint corresponds to an instant {\em snapshot} of the network, and the whole sequence becomes equivalent to the {\em evolving graph} model~\cite{Fer04}. 
At the other side of the spectrum, i.e. taking $\tau = \T$, the sequence would consist of a single footprint aggregating all interactions over the network lifetime.

\subsubsection{Indicators and Discussions}

We now discuss the definitions and peculiarities of a set of atemporal parameters, some of which are illustrated upon recent experimentations results (from~\cite{AQ2010a}) on the hep-th (High Energy Physics Theory) portion of the arXiv website. 
The dataset consists of a collection of papers and their related citations over the period from January 1992 to May 2003. For each paper the set of authors, the dates of on-line deposit, and the references to other papers are provided. There are 352 807 citations within the total amount of 29 555 papers written by 59 439 authors. From the dataset we extract the network of the most proficient authors - i.e., the authors of papers which received more than 150 citations. In all the example charts, a one-year time window is used.

\paragraph{Evolution of  the Density.}
One important and widely used indicator aimed at measuring the network structure is   
the density, which measures how close it is to a complete graph. The {\em density} of a graph $G=(V,E)$ is defined as: 
$$ D =  \frac {|E|}   {|V|  * (|V|-1)}$$ 

\begin{figure}[h!]
 \centering
 \includegraphics[width=52mm]{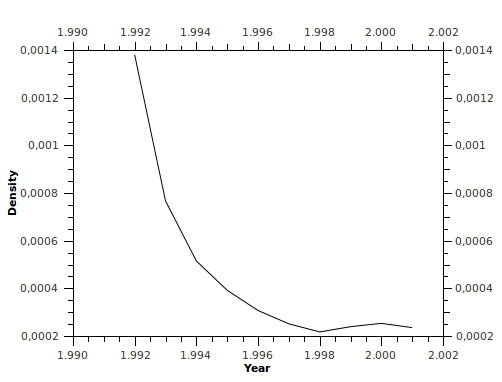}
 \caption{Evolution of the density.}
 \label{fig:density}
\end{figure}

The {\em evolution of the density} could be observed by looking at its trend over the sequence of footprints $SF=G^{\tau_1}, G^{\tau_2}, \dots, G^{\tau_i}$.
The trend of this value reflects the network's topology formation during time from a global perspective. It could be useful in many cases, such as in the study of transportation networks, e.g. to see how the equipment (number of roads, railways, flights connections...) increases over time. Figure~\ref{fig:density} provides another example showing a trend of {\em un-densification} observed in the above-mentioned scientific publishing network.

This counter-intuitive trend can be explained by an increasing number of authors. (Recall that these experimentations considered that the set of nodes in the footprint sequence was varying among the $G^{\tau_i}$s, based on the existence of adjacent edges in the considered footprints.)

\paragraph{Evolution of the Clustering Coefficient.}

The clustering coefficient is used in social network analysis to characterize architectural aspects. Several studies (e.g.,  \cite{holland1998,WATTS99}) suggest that in general nodes tend to create tightly compact groups characterized by a relatively high density of ties. Roughly speaking, the {\em clustering coefficient} of a node indicates how close to a clique its neighborhood is. It is formally defined in~\cite{WATTS99} as
 
$$C(x) =  \frac{|\{(u,v): u,v \in N(x)\}|} {deg(x)(deg(x)-1)}$$

The average clustering coefficient of a graph can then be defined as the average over all nodes: 

$$ AC= \frac{1}{|V|}  \sum_{x\in V}  C(x)$$ 

\begin{figure}[h!]
 \centering
 \includegraphics[width=50mm]{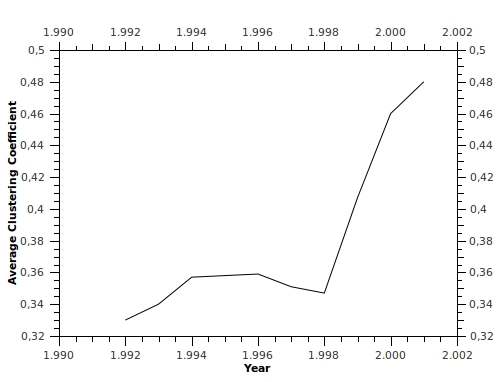}
 \caption{Average Clustering Coefficient Evolution}
 \label{fig:clustcoeff}
\end{figure}

As for the density, the evolution of these properties could be observed through measuring it on the footprints of SF. An increasing or decreasing trend of clustering coefficient would typically capture the formation or dismemberment of social communities at a global scale. An example is provided on Figure~\ref{fig:clustcoeff}, still with the same dataset, which shows that the connectivity first tends to be sparse, then after a phase transition around 1999, the nodes start to cluster into denser sub-communities.

\paragraph{Evolution of the Modularity.}

Modularity measures how the structure of a given  network is modular, {\it i.e.,} how it can be decomposed into subparts. It also quantifies the quality of a given network division into modules or communities. Networks with high values of modularity are characterized by dense {\em intra}-module connections and sparse {\em inter}-module connections. 

The {\em modularity} of a pair of nodes $u$ and $v$ is defined as

$$M(u,v) = \frac{deg(u)*deg(v)}{2|E|}$$

\begin{figure}[h]
 \centering
 \includegraphics[width=52mm]{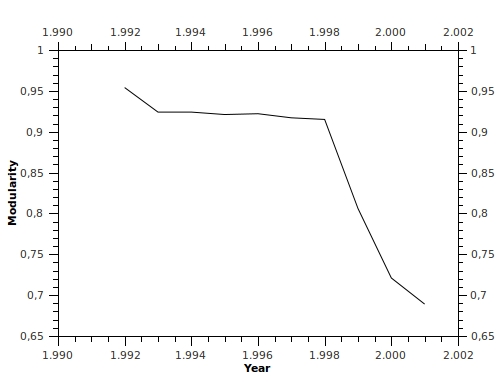}
 \caption{Evolution of the Modularity}
 \label{fig:modularity}
\end{figure}

The most common use of modularity is the detection of community structures (e.g. \cite{blondel08}). Such an indicator, if observed over time, can provide very interesting hints for the analysis of complex dynamic networks, in particular for the evolution of their structures and groups formation. It could also enable to see whether communities tend to specialize and/or homogenize. Figure~\ref{fig:modularity} shows the evolution of the {\em average} modularity over the sequence of footprints of our scientific networking example.

In a similar way as for the clustering coefficient, the evolution of modularity exhibits a phase transition around 1999 that separates a monotone trend from a decreasing one. This means that nodes first tends to form separate groups, which at some point start to interconnect with each other into a smaller number of larger groups (formation of communities). Modularity and clustering coefficient are clearly related. It was shown for example in~\cite{barmpoutis2010} that networks with the largest possible average clustering coefficient are found to have a {\em modular} structure, and at the same time, to have the smallest possible average distance between nodes.

\paragraph{Evolution of the Degree Power Law.}

Real world networks are ``scale-free'', in the sense that their node degree distributions follow a power-law that is not affected by the size of the network.
Such a power law indicates that the fraction $F$ of nodes that have degree $k$ decreases as $F(k) \sim k^{-\gamma}$, where $\gamma \in \mathbb{R}$ is a parameter that varies among different types of networks; its value is generally in the interval $[2,3]$.

\begin{figure}[h!]
 \centering
 \includegraphics[width=52mm]{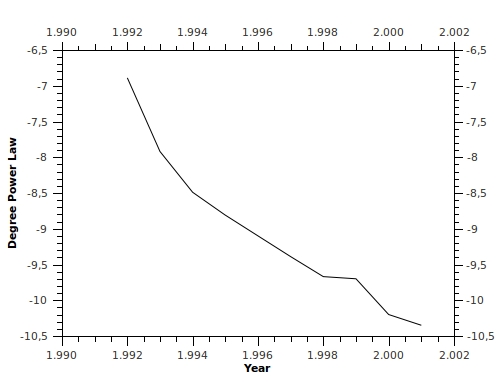}
 \caption{Evolution of the degree power law}
 \label{fig:degreepl}
\end{figure}

The evolution of the power law over time could reflect for example the arrival or departure of hubs -~nodes that interconnect several groups. Figure~\ref{fig:degreepl} shows the evolution of the power law exponent over the sequence of footprints of our dataset. As our example deals
 with the network of most proficient authors, i.e. a
subset of the dataset, the values in Fig \ref{fig:degreepl} are slightly
different from the traditional reference values.
In particular, the graphic shows  how closely the degree distribution of a graph 
follows a power-law scale at each time interval.
The higher the values, the more unequal is the distribution of connections within the nodes of the network.

Notice that the curve in Figure \ref{fig:degreepl} provides   additional details about the interaction pattern evolution of the network. As the evolution of the clustering coefficient shows an increase of the clustered structure of the network, and the modularity indicates that such an increase is characterized by the connection among separated groups, the decrease of the degree power law shows that the interconnection process is driven by nodes with low degree acting as hubs within groups.

\paragraph{Evolution of the Conductance.}

Social networks are intensively studied not only with respect to their structure, 
but also regarding the interactions occurring on top of them. For instance, 
several studies focused on {\em information diffusion} within groups based on a process of social influence ({\em influential networks}~\cite{Kats2005}). Such a process was intensively studied under the name of {\em viral marketing} (see for instance~\cite{Domingos2001})  to predict the propagation time of a message over a network. It was recently shown in~\cite{CLP10} that the {\em conductance} -~a measure that characterizes the time of convergence of a random walk toward its uniform distribution~- plays an important role in ``push-pull'' based dissemination strategies. The conductance of a graph is defined as the minimum conductance over all the possible cuts $(S,\bar S)$ in this graph (a {\em cut} is a partition of the nodes into two disjoint subsets). The conductance of a cut $(S,\bar S)$ is defined as 


$$ \varphi(S) = \frac{|(x_{\in S},y_{\in \bar S}) \in E|} {\min(|(x_{\in S},y_{\in V}) \in E|, |(x_{\in \bar S},y_{\in V}) \in E|)} $$

The evolution of conductance might reveal how the links of a network are organizing according to the distance between nodes, and indirectly reflect a process of self-optimization (or deterioration) of the network efficiency.

\subsection{Evolution of Temporal Indicators} 
\label{sec:temporal}

\subsubsection{Methodological approach}

Most temporal concepts --~including all those mentioned at the end of Section~\ref{sec:journeys}~-- are based on replacing the notion of {\em path} by that of {\em journey}. As a result, they can be declined into three versions depending on the type of distance considered ({\it i.e.,} shortest, foremost, fastest).  Since journeys are paths over  {\em time}, the  evolution of parameters based on journeys {\em cannot} be studied using a sequence of aggregated static graphs. For example,  there might be a path  between $x$ and $y$ in all footprints, and yet possibly no journey between them depending on the precise chronology of interaction. To analyze the evolution of such parameters, we need to use a more powerful tool: a sequence of time-varying graphs. 

\paragraph{TVGs as a sequence of (shorter) TVGs.}
Subgraphs of a time-varying graph \TVGL\
can be defined in a classical manner, by restricting the set of vertices or edges of $\G$. More interesting  is the possibility to define a \emph{temporal subgraph} by restricting 
the lifetime $\T$ of $\G$, leading to the graph $\G'=(V,E',\T',\rho')$ such that 
\begin{itemize}
\item $\T' \subseteq \T$
\item $E' = \{e \in E : \exists t \in \T' : \rho(e,t)=1\}$
\item  $\rho': E' \times \T' \rightarrow \{0,1\}$ where
$\rho'(e,t)=\rho(e,t)$ 
\end{itemize}
In the same way as for the sequence of footprints SF, we can now look at the evolution of a TVG through a sequence of shorter TVGs {\em ST}$(\tau) = {\cal G}^{\tau_0},{\cal G}^{\tau_1}, \ldots$, in which the intervals are either disjoint or overlapping.

\subsubsection{Indicators and Discussions}

\paragraph{Evolution of the (temporal) Distance.}
   
The basic concept of this class of indicators is that of {\em distance}. In particular, there are three different types of distances -~shortest, fastest, and foremost~- that are respectively noted $d(u,v)$,   $\temp(u,v)$, and $  \fast(u,v)$. As discussed in the introduction, these concepts of distance are central in various contexts and were recently subject to several studies. Algorithms to compute optimal journeys according to the three types of distances are available in~\cite{BFJ03}. ({\em Distributed} analogues of these algorithms were recently proposed in~\cite{CasFMS10} and~\cite{CasFMS11}.) Computing the distance gives an idea of how reachable the nodes are from each other, and thereby constitutes a general bound on dissemination speed.

A concept symmetric to the one of temporal distance is that of {\em temporal view}, introduced in \cite{KosKW08} in the context of social network analysis. The temporal view (or simply {\em view}) that a node $v$ has of another node $u$ at time $t$, denoted $\phi_{v,t}(u)$, is defined as the latest (i.e., largest) $t'\le t$ at which a message received by time $t$ at $v$ could have been emitted at $u$; that is, in the TVG formalism,

\begin{center}
  \small
  $\max \{departure(\J) : \J \in\J^{*}(u,v) \wedge arrival(\J)\le t\}$.
\end{center}

This concept could, as that of distance, be declined into three versions (the above one is symmetric to the {\em foremost} distance). Studying the evolution of temporal distances or views over a sequence of temporal subgraphs reflects how close in time, or in hops, the nodes tends to become. It serves as a basis to most of the indicators discussed below.

\paragraph{Evolution of the (temporal) Diameter and Eccentricity.}

The three journey-based versions of eccentricity and diameter were first discussed in a communication network context~\cite{BFJ03}. The eccentricity of a node $u$ in a TVG $\G$ can be defined in terms of {\em shortest} journeys as
$$ e(u) =  \max \{  d(u,v): v \in V  \}$$
where    $d$ can be substituted by $\temp(u,v)$ or 
$\fast(u,v)$  to obtain the foremost eccentricity  ${\ecc}(u) $, 
or  the fastest eccentricity   $\hat{\ecc}(u)$,  respectively.
The eccentricity of a node directly reflects its reachability capacity, and therefore the impact it can have on the network. Such a parameter could have a particular significance in some field of research, e.g. in epidemics, the existence of nodes with a high temporal eccentricity could be associated with the possibility for a virus to survive long-enough to reinfect people. Three versions of the diameter naturally follow based on those of eccentricities: $max\{e_i(u):u\in V\}$, 
$max\{{\ecc}_i(u):u\in V\}$,  and 
$max\{\hat{\ecc}_i(u):u\in V\}$. 
The foremost version of the temporal diameter was specifically studied in~\cite{ChMMD08} from a stochastic point of view by Chaintreau {\em et al.}, but to the best of our knowledge, the {\em evolution} of the temporal diameter or eccentricities have never been considered yet. Looking at them could reveal complex social parameters, e.g., considering the evolution of standard deviations among node eccentricities could reflect how a network tends to create {\em fairness} or {\em inequalities} among its participants.

\paragraph{Evolution of the (temporal) Centrality.}

One of the most important properties of social networks' structures is the so-called notion of {\em power}. As a shared definition of power is still object of debate, the design of metrics able to characterize its causes and consequences is a pressing challenge. In particular the social network approach emphasizes the concept of power as inherently relational, {\it i.e.,} determined by the network topology. Hence, the focus must be put on the relative positions of nodes. In order to characterize such a property the concept of {\em centrality} has emerged. The simplest centrality metric, namely the {\em degree centrality}, measures the number of edges that connect a node to other nodes in a network. Over the years many more complex centrality metrics have been proposed and studied, including {\em Katz status score}  \cite{katz53}, {\em $\alpha$-centrality}  \cite{Bonacich2001}, {\em betweenness centrality}  \cite{Freeman1979}, and several others based on random walk~\cite{Dong2002,Stephenson1989}, the most famous of which is the {\em eigenvector centrality} used by Google's PageRank algorithm~\cite{Page99}. 
The temporal adaptation of these concepts is meaningful, and Kleinberg et al. have shown in~\cite{KosKW08} that nodes that are topologically more central are not necessarily central from a temporal point of view, whence the concept of {\em temporal centrality}. Studying the evolution of these over time could in turn shed light on how ``powerful'' nodes tends to emerge in a network. {\em Betweenness} and {\em closeness} are two well-known measures of centralities.\medskip

\noindent{\em Temporal betweenness.}
The betweenness of a node in a static graph 
measures the occurrences of that node within the shortest paths of other nodes~\cite{Freeman1979}. A temporal version of the betweeness based on foremost journeys was considered in recent work by Tang {\em et al.}~\cite{TMMLN10}. The definition can be generally formulated as
$$    B(q) = \sum_{v \neq u \neq q \in V} \frac{|d'(u,v,q)|}{|d(u,v)|}$$
where  $|d(u,v)| $ is the number of shortest 
journeys between $u$ and $v$  in the time varying graph $\G^{\tau_i}$, and 
$|d'(u,v,q)|$  is the number of shortest journeys, among them, that pass 
through $q$. We can analogously define  the    temporal betweeness 
 in terms of foremost or fastest  distance, by substituting  $d(u,v)$ with 
 $\temp(u,v)$ or  $\fast(u,v)$.\medskip
 
\noindent{\em Temporal closeness.}
In a static context, the closeness measures the mean of the shortest paths between a node and all the other reachable nodes~\cite{Freeman1979}. It can be formally defined as
$$ TC(u) =  \sum_{v \in V \setminus u}   \frac{d(u,v)}    {|\{w\in V : \exists \J\in \J^*(u,w)\}|} $$
and again, possibly declined to a shortest, foremost ($\temp(u,v)$), or fastest ($\fast(u,v)$) versions. As one will certainly notice, this parameter is highly related to that of temporal eccentricity, and yet, both have appeared in very different fields of research. This illustrates again how general both the temporal concepts and the formal tools can be.

 \bibliographystyle{plain}

\end{document}